\documentclass[preprint]{aastex7} 
\usepackage{soul}

\newcommand{\alf}{Alfv\'en}
\newcommand{\sm}{$\sim$}
\newcommand{\kms}{km~s$^{-1}$}
\newcommand{\ha}{H$\alpha$}
\newcommand{\CaII}{Ca {\sc ii}}
\newcommand{\RBE}{blueshifted spicules}
\newcommand{\RRE}{redshifted spicules}

\begin{document}


\title{Spatial and Dynamical Relations between Spicules and Network Bright Points}

\author[0000-0002-5865-7924]{Jeongwoo Lee} 
\affiliation{Institute for Space Weather Sciences, New Jersey Institute of Technology, Newark, NJ 07102, USA}
\affiliation{Center for Solar-Terrestrial Research, New Jersey Institute of Technology, Newark, NJ 07102, USA}
\affiliation{Big Bear Solar Observatory, New Jersey Institute of Technology, Big Bear City, CA 92314, USA}
\email{leej@njit.edu}

\author[0000-0002-7358-9827]{Eun-Kyung Lim}
\affil{Korea Astronomy and Space Science Institute, Daedeokdae-ro, Yuseong-gu, Daejeon 34055, Republic of Korea}
\affil{Astronomy and Space Science, University of Science and Technology, 217 Gajeong-ro, Yuseong-gu, Daejeon 34113,
Republic of Korea}
\email{eklim@kasi.re.kr}

\author[0000-0003-0975-6659]{Viggo Hansteen}
\affiliation{SETI Institute, 339 Bernardo Ave, Mountain View, CA 94043, USA}
\affiliation{Lockheed Martin Solar and Astrophysics Laboratory, 3251 Hanover Street, Building 203, Palo Alto, CA 94306, USA}
\email{vhansteen@seti.org}
\affiliation{Rosseland Centre for Solar Physics, University of Oslo,  PO Box 1029 Blindern, 0315 Oslo, Norway}

\begin{abstract}
Spicules are among the most ubiquitous small-scale, jet-like features in the solar chromosphere and are widely believed to play a significant role in transporting mass and energy into the solar corona with their mechanisms not fully understood. We utilize high-resolution \ha\ images acquired from the 1.6-meter Goode Solar Telescope (GST) at Big Bear Solar Observatory (BBSO) to investigate spatial and the dynamical properties of both spicules and network bright points (NBPs) and, for the first time, incorporated NBP motions in the analyses of spicules. Our main results are as follows: (1) The speed distributions of blueshifted spicules and NBPs both exhibit distinct peaks, whereas that of redshifted spicules is monotonically decreasing. (2) Torsional motions of spicules inferred from alternating signs of Dopplershifts are faster than the NBPs' transversal motions by a factor of 10--$10^2$, which may imply the mass density ratio in two different heights as $10^2$--$10^4$. (3) Blueshifted spicules are found to be more abundant than redshifted spicules in general, but their relative population difference reduces to \sm10\% at Doppler speeds above \sm35 km s$^{-1}$. (4) Redshifted spicules lying at higher heights share morphological and dynamical similarity with the blueshifted spicules, which implies the same driving mechanism operating in both directions. (5) These two populations appear above NBPs concentrated under the AIA 193 \AA\ bright region. We interpret these results in favor of a scenario that \alf\ waves generated by NBPs motions impart their energies to spicules in both torsional and field-aligned motions, and also contribute to the coronal heating and possibly the acceleration of the solar wind.
\end{abstract}


\section{Introduction} \label{sec:intro}
In high-resolution solar images, the quiet sun is active in a variety of forms of small-scale jets, which may provide the upward flux of mass, momentum, and energy necessary for the coronal heating and acceleration of the solar wind. 
Improved physical understanding of the formation of and the mechanisms behind the jetting phenomenon is fundamentally important within the broad field of the solar physics and helio physics.
Solar spicules as the smallest class of such solar ejections are abundant vertical ``spikes'' seen across the solar disk, including coronal holes \citep[e.g.,][]{beckers1968}. 
Using high-resolution, high-cadence solar limb observations in the \CaII\ H (3968~\AA) line obtained by Hinode's Solar Optical Telescope \citep[SOT;][]{Tsuneta+etal2008SoPh..249..167T}, two kinds of spicules were defined. 
Type I or ``classical'' spicules have a lifetime of 3--7 minutes and display upward plasma flows (15--40~\kms) usually followed by downward returning flows, and barely penetrate into the corona  \citep{DePontieu+etal2007ApJ...655..624D}. Type II spicules are not followed by downward motions but only faded out, implying potential role in both coronal heating and mass supply \citep{DePontieu+etal2007PASJ...59S.655D}. They are dominated by high speed (30--110~\kms) upward flows, rapid heating with short life times (10--150~s), and large vertical extent (up to 7 Mm) dominating the interface between the chromosphere and the corona \citep{Pereira+etal2012ApJ...759...18P, Pereira2016}. 

While spicules refer to thin structures observed at the limb, ground-based data revealed that their probable on-disk counterparts do appear jet-like and show dynamic behavior, sometimes spinning or splitting over magnetic bases \citep{Suematsu2008ASPC..397...27S, Samanta2019}. 
The disk counterpart of Type II spicules in the form of significant absorption in the blue wing of the chromospheric line profiles of H$\alpha$  was found and termed ``upflow events'' 
\citep{WangH+etal1998SoPh..178...55W,Chae+etal1998ApJ...504L.123C,LeeCY+etal2000ApJ...545.1124L}
or rapid blueshifted excursions (RBEs) defined using Ca {\sc ii} 8542 lines \citep{Langangen+etal2008ApJ...679L.167L}.
These upflowing spicules are regarded as the on-disk counterpart of type II spicules \citep{Rouppe+etal2009ApJ...705..272R} and are expected to contribute to the corona either in energy or in mass.
Many studies have explored the connection between the chromosphere and the corona in the context of spicules—particularly investigating whether plasma in Type II spicules is heated to coronal temperatures and whether these upward-moving spicules contribute significantly to the mass of solar corona \citep{DePontieu+etal2009ApJ...701L...1D, DePontieu+etal2011Sci...331...55D, 
DePontieu2017ApJ...845L..18D,
Judge2012ApJ...746..158J,
Henriques2016ApJ...820..124H}. 
This topic has remained an active area of research in recent years 
\citep{Bose2023ApJ...944..171B, Bose2025ApJ...983L...7B}.
It is, however, not yet fully established whether Type II spicules are globally a significant source of energy in the corona  \citep{Klimchuk2012, Klimchuk2014, Sow2022}.
 
Although the red wing counterparts of RBEs in the {\ha} and {\CaII} 8542 {\AA} lines were found and termed rapid red shifted excursions \citep[RREs;][] {Sekse+etal2013ApJ...769...44S}, they were not immediately identified as the on-disk counterpart of the downflowing spicules. Their redshifts were rather considered as due to either transverse motion, swaying motion, torsional motion, or {\alf} waves under geometric effects \citep{Sekse+etal2013ApJ...764..164S, Kuridze2015}. Only later were rapid spicular downflows in the {\ha} line moving along the chromospheric field in the disk confirmed and termed downflowing RREs \citep[dRREs;][]{bose2021a}.
Three types of spicular motions are established using observations with Swedish 1 m Solar Telescope (SST): field-aligned flows of order 50--100 \kms, swaying motions of order 15--20 \kms, and torsional motions of order 25--30 \kms\ \citep{DePontieu12}. 
The coexistence of these longitudinal and transversal motions makes it challenging to reliably distinguish between upflowing and downflowing spicules, particularly in on-disk observations \citep{ Sekse+etal2013ApJ...769...44S}. 

Found under spicules are the network bright points  \citep[NBPs;][]{2000Muller}, which always reside inside magnetic fields and are also called magnetic bright points. Earlier studies suggested that
their unusual brightness be associated with them being footpoints of open field lines \citep{2000Muller, Spruit+Scharmer2006A&A...447..343S}. \citet{dunn+zirker1973} suggested that a spicule should be related to granulation based on the similarity between the birth rate of spicules and that of granules \citep{Becker1964}. 
The idea on the generation of spicules and waves by NBPs' motions also dates back to \citet{dunn+zirker1973} and further studied by \citet{Lee2025ApJ...988L..16L} until recently. 
As a comparison, the energy of individual {\alf}ic pulses generated by NBPs is estimated to reach \sm$10^{25}$ erg \citep{Lee2025ApJ...988L..16L}, and thus carrying sufficient amount of energy for coronal heating
\citep{McIntosh2011}.
Therefore, two distinct perspectives can exist on addressing the coronal heating problem. One posits that the primary energy input originates in the chromosphere, where localized heating processes drive energy upward \citep{2001Aschwan}. The other suggests that energy is transported directly from the photosphere and dissipated in the corona through mechanisms such as stochastic topological reconnection or braiding of magnetic field lines \citep[cf.][]{Parker1988}.

This study aims to extend previous investigations of solar spicules by incorporating network bright points (NBPs) into the analysis of the spicule–corona connection based on high-resolution \ha\ images acquired from the 1.6-meter Goode Solar Telescope (GST) at Big Bear Solar Observatory (BBSO).  The target region is positioned approximately midway between the solar disk center and the limb, enabling simultaneous observation of both spicules and NBPs. In this geometry, however, the measured Doppler velocity of a spicule reflects a combination of longitudinal (field-aligned) and transverse (swaying or torsional) motions.
Accurate characterization of both swaying motions and rapid field-aligned flows requires either a sufficiently large field of view (FOV) or high-cadence imaging. Paired redshifted/blueshifted patterns appearing at adjacent spatial locations can, a priori, be regarded as indicative of torsional motions. We therefore focus on their association with NBPs and the corresponding EUV emissions. We avoid the conventional terms rapid blue-shifted excursions (RBEs) and rapid red-shifted excursions (RREs) in order to encompass the full velocity range of these features. Instead of up or downflows, we adopt the more general notations, \RBE\ and \RRE\ to describe the spicules under study.

\begin{figure*}[tbh]  
\includegraphics[width=\textwidth]{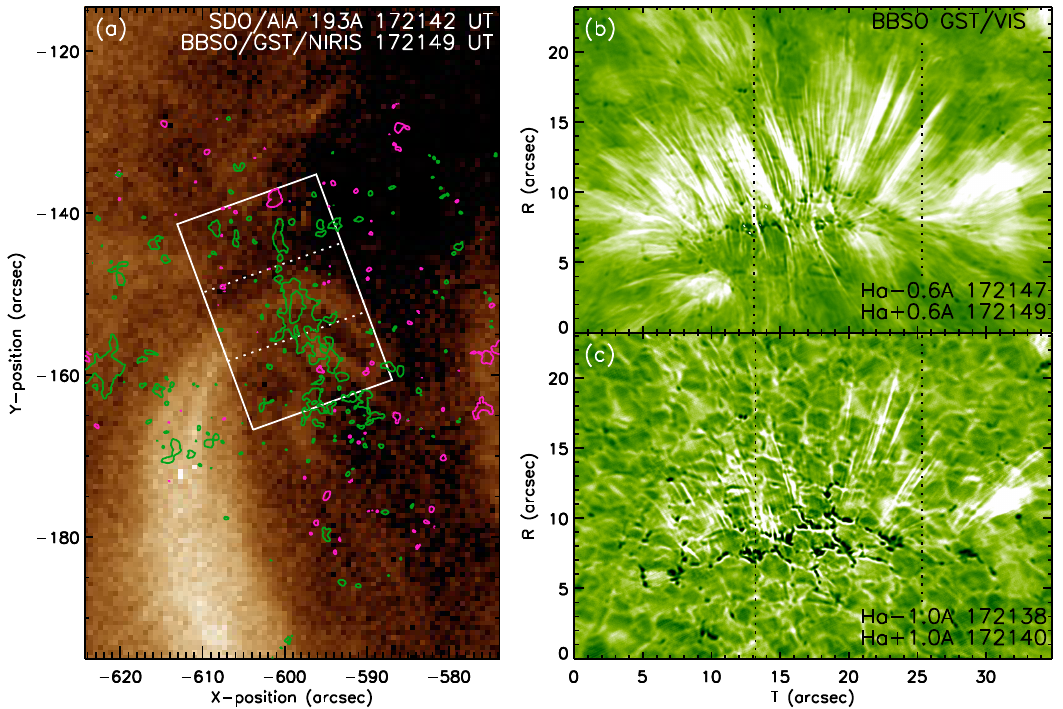}
\caption{EUV and {\ha} images. (a) SDO/AIA EUV 193 {\AA} shows the target region near a coronal hole. The NIRIS magnetogram is overplotted as contours at the levels of +50 G (pink) and $-$50 G (green). The white box is plotted to denote the FOV of the other panels, and it has the ordinate in the solar radius direction (named $R$) and the abscissa in the azimuthal direction ($T$).
The rest are GST/VIS {\ha} composite images constructed by adding the blue/red wing images in the {\ha}$\pm$0.6 {\AA} (b) and those in $\pm$1.0 {\AA} (c). Both images are inverted in intensity so that the bright straw-like features are spicules and the dark points underneath are NBPs. }
\label{f1}
\end{figure*}

\section{Data and Methods}

On 2018 July 29 the GST was targeting a small region near the magnetic field network boundary region near a disk coronal hole.
Installed on the BBSO/GST are the Visible Imaging Spectrometer \citep[VIS;][]{Cao+etal2010SPIE.7735E..5VC} and magnetograms from Near-infrared Imaging Spectropolarimeter \citep[NIRIS;][]{Cao2012}. From these instruments we obtained high spatial resolution (0.24$''$) magnetograms and high resolution (0.10$''$) {\ha} multi-wavelength images. 
According to EUV images from Solar Dynamics Observatory Atmospheric Imaging Assembly \citep[SDO AIA;][]{Lemen+etal2012SoPh..275...17L, Pesnell.SDO.2012SoPh..275....3P} our target region lies in the boundary of a coronal hole located about a half radius from the disk center. 
No large-scale events occurred near this region during the observation such as flares and coronal mass ejections. This dataset was used elsewhere for different purposes: magnetic cancellation \citep{WangJ2022}, magnetic driver of the solar wind \citep{Raouafi2023}, driver for switchbacks \citep{Leej2024}, potential energy calculation \citep{Manolis25}, and wave generation \citep{LeeJ2025ApJ...988L..16L}. The focus of this study is the spicules and the NBPs.

At the time of this observation, the instrument was set up to produce high-resolution images at 11 wavelength points  {\ha}$\pm$0.2$,\pm$0.4, $\pm$0.6, $\pm$0.8, $\pm$1.0 {\AA} and the line center. In the VIS data, a pair of blue and red wing images are separated by only 2 s, whereas a set of the 11 wavelength images is separated from the next one by 40 s. 
We use these 11-wavelength point images from the Fabry–Pérot interferometer to construct Dopplergrams point by point basis. 
In each wavelength, the images are normalized by the exposure time, and the average profile over the whole FOV is calibrated to match the standard solar {\ha} profiles from the Ground-Based Solar Observations Database BASS 2000 \citep{Paletou2009}.

We first  use so-called Optical Flow method to co-align the line center image with the off-band images \citep{Cai2022, 2022XYang}.
Using this co-aligned image set, we calculated, on a point by point basis, the center-of-gravity (COG) wavelength, $\lambda_{\rm COG}$, which is defined by the local residual intensities between the observed $I_\lambda$ and the reference profile $I_{\rm ref}$.
Two types of $\lambda_{\rm COG}$ may result, depending on which profile to use as $I_{\rm ref}$: the ambient continuum intensity \citep{Uitenbroek2003} or the mean spectrum of the entire FOV \citep{Rouppe+etal2009ApJ...705..272R}. 
These two methods called methods 1 and 2, respectively, produce differing results aside from the magnitudes of speeds. In Appendix \ref{ap:1} we tested an alternative method (method 3) in which method 2 is modified to be consistent with method 1 except the magnitude. We adopted method 1 to produce stable Dopplergrams at the cost of denouncing the high speeds, and method 3 when high speeds of {\RBE}/{\RRE} are needed to be determined. Conversion between these two types of speeds is available through an empirical relationship as demonstrated in Appendix \ref{ap:1}.
We call this Dopplergram a pseudo-Dopplergram in the sense that it correctly detects the signs of Doppler signals but may underestimate their magnitudes. Not that 11 wavelength points are too few, but that the spectral window, {\ha}$\pm$1.0 {\AA}, is not always wide enough to fully determine the LOS speed of fast spicules. 

To trace the subtle motions of NBPs,
we use the Southwest Automatic Magnetic Identification Suite \citep[SWAMIS;][]{DeForest_2007}. 
SWAMIS was developed for tracking time-dependent features in magnetograms and can handle both positive and negative signals near noise levels. We here use SWAMIS to trace NBPs on H$\alpha$ filtergrams by setting a threshold in the {\ha} intensity contrast, $(I-I_b)/I_b \geq 7$\% for qualifying NBPs.
The feature recognition algorithm requires discrimination to separate the foreground features from the background noise.
We track irregularly shaped NBPs and represent them as a circle centered on the center-of-mass position of each mask with equivalent diameters. 
We measure the displacement of the masks with the same ID over consecutive frames, and convert them to the velocities using ${\bf v}^k = ({\bf x}^{k+1}-{\bf x}^k)/\Delta t$ with the frame-to-frame time interval $\Delta t$ typically 40 s. 

Figure \ref{f1} shows the spicules and NBPs in comparison with the EUV image. 
Figure \ref{f1}a displays the EUV images from SDO/AIA 193 {\AA} data, which is dominated by plasma around 1.5 MK. The overplotted contours are from
the NIRIS magnetogram contours. The dark feature is the coronal hole on the disk and low latitude. The darkness arises as due to escaping material along open fields \citep{Pneuman1973}. The over-plotted contours represent the line-of-sight (LOS) magnetic fields at +50 G (pink) and $-$50 G (green) from the NIRIS magnetogram, and the white box denotes the FOV of the other panels (Figure \ref{f1}bc). 
The NIRIS magnetogram confirms that the bright EUV loop-like structure arises from the negative magnetic polarity only. 

Figure \ref{f1}bc are ``composite'' {\ha}  wing images obtained by adding a pair of red and blue wing images at $\pm$0.6 \AA\ (b) and $\pm$1.0 \AA\ (c).
Spicules appear as dark straw-like features  \citep{2000Sterling}, which are, in these inverted images, bright slender features. In the wavelengths farther from the {\ha} line center the spicules are less populating and more sparsely spaced, because only those with higher Doppler speed can be detected. 
We identify those spicules in the far wing $\mp$1.0 {\AA} (Figure \ref{f1}c) with blueshifted/{\RRE}, respectively.
The far wing images also show the granules and NBPs more clearly. 
NBPs are regarded as footpoints of open field lines, and appear bright (looking dark in these inverted {\ha} images) as the opacity drops due to material escape along the open field lines \citep{Spruit+Scharmer2006A&A...447..343S}. 
It is the main task of this study to investigate which one among blueshifted/{\RRE} and NBPs better align with the coronal brightness.

\begin{figure}[tbh]  
\includegraphics[width=\textwidth]{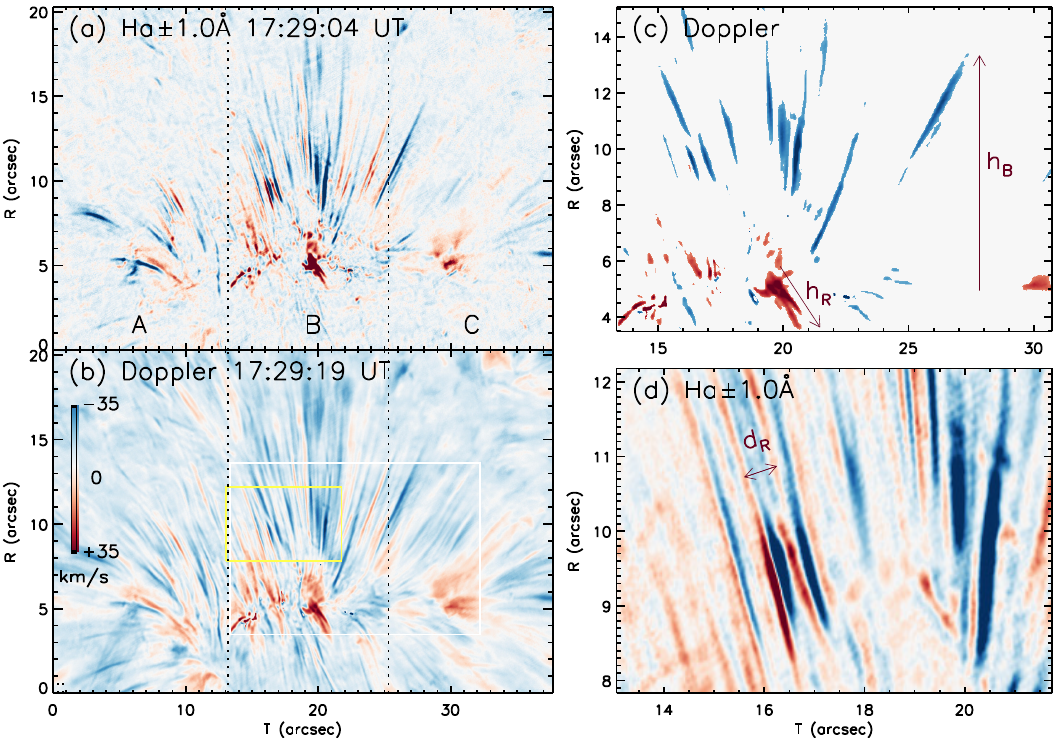}
\caption{Blueshifted and \RRE\ maps and NBPs. 
(a) A wing difference image obtained by subtracting {\ha}$-1.0${\AA} and {\ha}$+1.0${\AA} images.  
(b) A pseudo-Dopplergram of the {\ha} line constructed using images in the 11 wavelengths between {\ha}$\pm1.0${\AA} with a scale bar.
The vertical dotted lines divide the FOV into 3 sections, A--C.
(c) The sub-region within the white box in (b) is shown with only Doppler signals above 25 {\kms}. The apparent height of a tall {\RBE} and a {\RRE} are $h_B\approx$8.4$''$ and $h_R\approx$2.7$''$.
(d) Another sub-region, the yellow box in (b), is shown to be filled with thin and long {\RRE} and {\RBE}.
$d_R\approx$0.65$''$ is the interval between two adjacent thin {\RRE}.
}
\label{f2}
\end{figure}

\section{Results} 

We first sort out types of spicular motions in \S 3.1. We investigate spatial correlation of spicules and NBPs in \S 3.2, and their speed distributions in \S 3.3. We present time-distance (TD) maps for spicules and NBPs in \S 3.4, and finally compare them with the coronal emission in \S 3.5.

\subsection{Types of Spicular Motions}

Figure  \ref{f2} shows wing-difference images and Pseudo-Dopplergrams for checking the nature of spicule motions.
The far wing difference image (Figure \ref{f2}a) is created by subtracting the blue wing image from the red wing image, and plotted in a color scheme, which distinguishes {\RBE}s (blue) from {\RRE}s (red). 
Figure \ref{f2}bc are the pseudo-Dopplergram constructed in this way, for comparison with the wing difference images (Figure \ref{f2}ad). They agree to each other very well and justify our further use of the pseudo-Dopplergrams.
The fine structures are best seen sharply in the wing difference image (a,d), and tend to be smoothed out after in the Dopplergram (b,c).
NBPs do not appear clearly in the wing difference images because their Doppler signals are weak. We therefore focus on their transverse motions calculated using SWAMIS on their images.
If we select only strong signals, the \RBE\ dominate mostly at greater heights, while the redshift signals lie in the lower heights. We can imagine that the former is due to an upward directed force while the latter may indicate gravitational downfall.
If we lower the threshold, the thin \RRE\ appear between {\RBE}, 
to produce many pairs of Doppler signals in both signs.

In the sub-region marked with the white box (Figure \ref{f2}c), weak Doppler signals below 25 {\kms} are removed in order to show high speed spicules only. In this case, the \RBE\ and \RRE\ are clearly separated in height, and the areas occupied by {\RBE} and {\RRE} are comparable.
Specifically, the height scales denoted as $h_B$ and $h_R$ are $6.1\times 10^3$ km and $1.9\times 10^3$ km, respectively, in the image. While their actual heights may differ due to projection, 
two types of spicules are clearly distinguished by height alone.
One is the thick and low-lying {\RRE} represented by $h_R$. This is one of the few examples which can be identified with dRREs representing materials falling under the gravity along the field lines. The other is the high-lying {\RBE}s represented by $h_B$, which is probably driven by additional force acting upward. 

If we lower the threshold, however, thin and long structures in redshift appear accompanying or surrounding the {\RBE}  (Figure \ref{f2}d). 
This speed dependent trend means that there is torsional motion in addition to the field aligned upward motion.  
Identification of torsional motions is generally less ambiguous in near-limb observations. In the present case, however, the Doppler signatures likely contain contributions from the three types of motions: field-aligned, swaying and torsional components. If the blue- and redshifted features are located very close to each other, they can a priori be regarded as a paired signature of torsional motion within a single spicule. In Figure \ref{f2}d, the separation between two redshifted spicules, $d_R/2$, 
is approximately 230 km, and each redshifted spicule has a width of about 100 km, suggesting that at least three spicules exhibiting torsional motions are present within the field of view. Depending on how the red- and blueshifted features are paired, the inferred torsional sense can be either entirely clockwise or include one counterclockwise case.
Otherwise, Doppler signals result from the superposition of torsional and swaying motions along the line of sight precluding straightforward interpretation \citep{Sekse+etal2013ApJ...769...44S}. If the single signed Doppler shifts are not torsional in origin, the very thin and elongated morphology of these blue and redshifted spicules could indicate a common physical mechanism capable of driving motions in both directions along a field line.

\begin{figure}[tbh]  
\includegraphics[width=\textwidth]{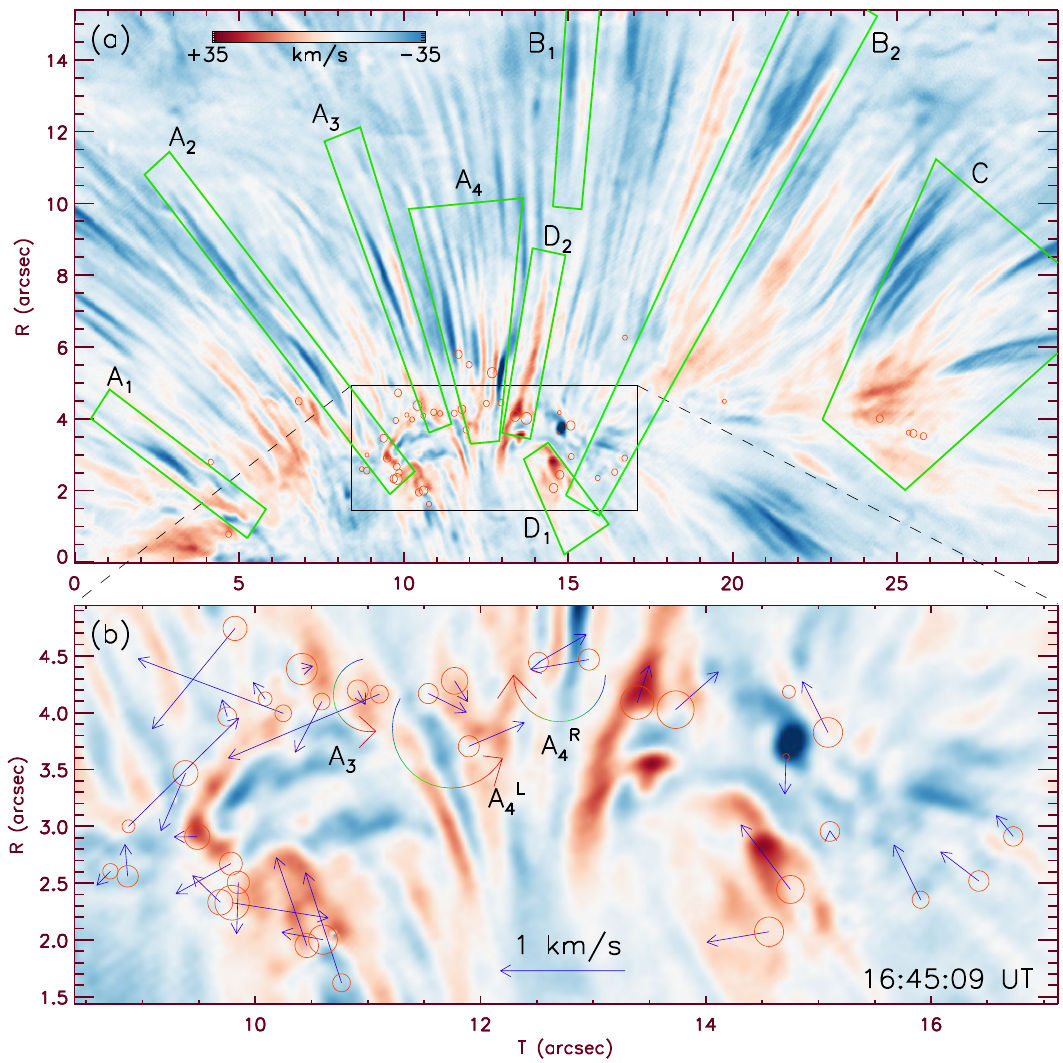}
\caption{Spicules and NBPs in motion.
Both panels use the Dopplergrams from the GST/VIS \ha\ line data at 16:45:09 UT a the background image. (a) Spicules of interest and seemingly associated NBPs are marked with the green trapeziums. The red circles mark the location of active (moving) NBPs with area-equivalent diameter. 
(b) Blue arrows are added representing the velocity of NBPs determined using SWAMIS. The three half circles with arrowhead represents inferred rotational sense of NBPs color-coded to mimic blue-red shifts. The half circle denoted A$_3$ is the inferred rotational motion of the footpoint of the spicule in box A$_3$, and those denoted A$_4^{\rm L,R}$, the leftmost and rightmost spicules in box A$_4$.
}
\label{f4a}
\end{figure}

\subsection{NBPs and Doppler Motions of Spicules}\label{sec:new}

We further investigate whether these torsional motions are correlated with the motions of nearby NBPs. 
In Figure \ref{f4a} we plot the NBPs over the Dopplergrams by the red circles with their position (circle center), equivalent area (circle diameter) and displacement (arrows) determined by applying SWAMIS to the \ha\ blue wing images. The green trapeziums mark spicules of interest and the NBPs potentially associated with the spicules in view of their trajectories.  

We classified the observed spicules into four distinct types.
Type A spicules exhibit thin, elongated structures with closely spaced blue- and redshifted components, suggesting they belong to a single flux tube.
Type B spicules share similar characteristics but are located at higher altitudes, showing weak or no Doppler signals at lower heights. This may indicate that their torsional motion is initially weak and becomes more pronounced as they propagate upward.
Spicules C are thick and diffuse, implying different mechanism from A or B type spicules. They exhibit Doppler signals in one sign all the way up or joined by the other sign above some height.  
Spicules D are redshifted with their downward motion identified from visual inspection, namely, dRRE.

There are also many thin spicules of single sign Doppler signals. We are not sure if they have no torsional motions or have torsional motions but concealed by the swaying motion \citep{Sekse+etal2013ApJ...769...44S}.
These spicules along with thin alternately signed Doppler signals tend to have strong NBPs underneath.
On the other hand, the thick features tend to have weak or no NBPs underneath. For instance, the only redshifted spicules as in the box C generally show no obvious correlation with NBPs. It may be that downward motions are neither torsional, nor associated with NBPs as drivers.

Torsional motions are most suspected for A and B types. It would be ideal if we could compare the sense of torsional motion with that of the NBP motions.
Unfortunately, we could measure only rectilinear translational motion of NBPs using SWAMIS, while NBPs may exhibit expansion, shrinkage, or deformation of the morphology \citep{Sharma2023NatAs...7.1301S, Wedemeyer-Bohm2012Natur.486..505W}. 
Nevertheless we attempt to identify the senses of rotation of NBPs
shown in Figure \ref{f4a}b by matching them to the spicules in Figure \ref{f4a}a.
For instance,  we infer anticlockwise and clockwise rotations of NBPs as denoted by the half circles with arrowheads, A$_4^{\rm L,R}$ in (b), which seem to match the Doppler sign pairs of the leftmost and rightmost spicules in box A$_4$ in (a).
However, the rotation of NBPs denoted by the half circle A$_3$ is not clear because one NBP exhibits a dominant speed. At least we can see a spicule with strong Doppler signal having prominent NBPs near its footpoint.
The spicule(s) in box A$_2$ has not only highly varying Doppler signal along its length but a complex distribution of NBPs underneath. Either the torsional wave has a relatively short wavelength or complex motions of NBPs affects the wave propagation.

The sense of torsional rotation does not seem to be uniform across the entire network area. The box denoted A$_4$ in Figure \ref{f4a}a contains at least 3 pairs of blue/redshifted spicules with different sense of rotation. Occasionally, a group of NBPs move around a local point in the same sense, in which case the torsional motion of the connected spicules may share the same sense of rotation as shown in Figure \ref{f2}d. 
The relationship between NBPs and spicules is also not strictly one-to-one, despite their general spatial association. In some regions, we observe more NBPs than spicules than NBPs (e.g., A$_2$), while in other regions the opposite is true  (e.g., A$_4$). The opacity gaps between spicules and NBPs further complicate efforts to establish direct connections between these two features. Such poor one-to-one correspondence is expected for several reasons, including time delays due to finite propagation speeds and the possible obscuration of NBPs by overlying spicules. Therefore, instead of seeking individual pairings, it would be more appropriate to examine the statistical relationship between NBPs and spicules.

\begin{figure}[tbh]  
\includegraphics[width=\textwidth]{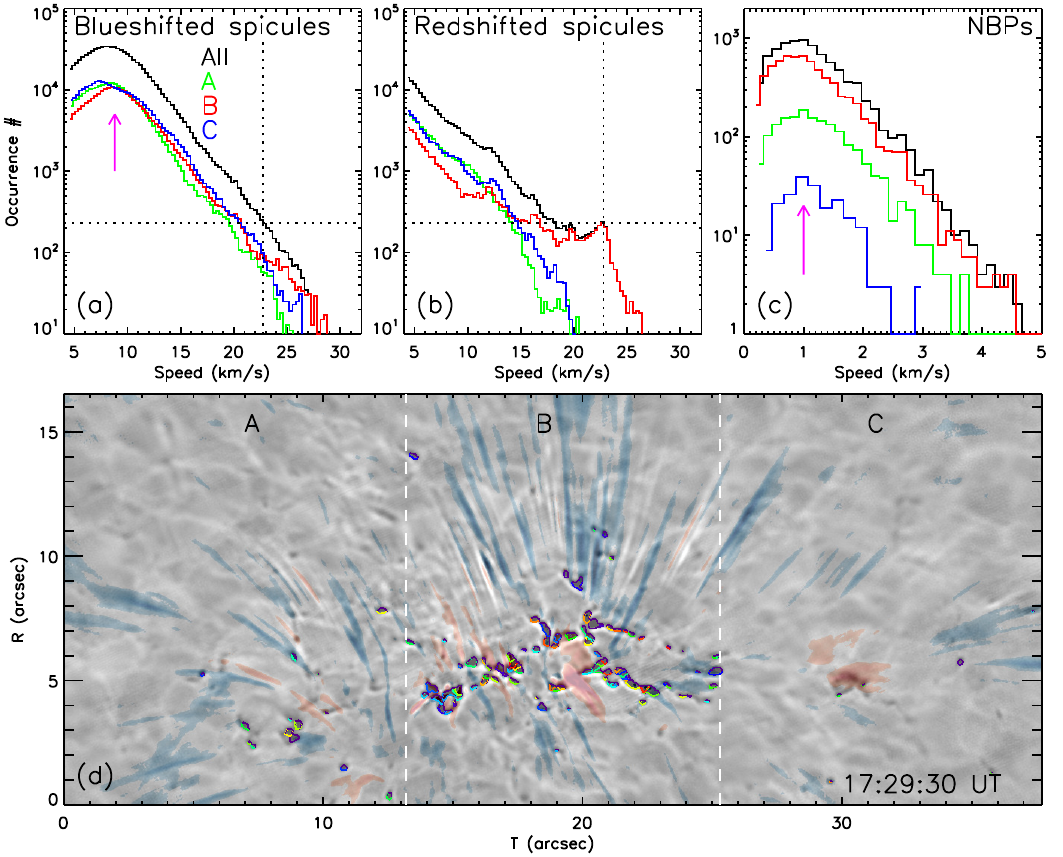}
\caption{Speed distributions in the chromosphere and photosphere.
The number of pixels per speed interval counted for (a) {\RBE}, (b) {\RRE}, and (c) NBPs. The dotted lines in (a) and (b) are identical and indicate that in the particular speed interval and region, the count of {\RRE} pixels agrees to that of {\RBE}, while {\RBE} usually outnumber {\RRE} in other speeds.
The speed distribution of NBPs in (c) is constructed from 3429 data points from total 141 frames.
(d) is the frame used in the calculation of the speed distributions in (a,b), in which the blueshifted/{\RRE} are marked in the blue/red color shades, and NBPs, in mixed colored contours. The vertical dashed lines divide the whole region into 3 sub-regions denoted as A--C.
}
\label{f4}
\end{figure}

\subsection{Speed Distributions of Spicules and NBPs}\label{sec:statis}

Having identified different types of spicules based on their heights and morphologies, we also found their relative occurrence varies with velocity. Figure \ref{f4}a--c show the speed distributions of the blue- and redshifted spicules, as well as that of the NBPs, as calculated from the image in Figure \ref{f4}d. There the spicules are colored according to the Doppler signs and NBPs are marked in mixed color contours. 
A glance at the bottom panel yields the impression that the {\RBE} dominate the {\RRE} in number and total area. 
For {\RBE}/{\RRE}, we counted the number of pixels in the single snapshot of pseudo-Dopplergram that lie in each speed interval (Figure \ref{f4}ab). 
The projection effects are  not removed, and the actual speeds of spicules must be higher. Spicules moving fast but without the characteristic of ``excursion’’ are also counted in.
The speed distribution of NBPs is constructed after tracing 3429 features from total 141 frames of the {\ha} far red wing image using the SWAMIS and shown in Figure \ref{f4}c.

However, the relative dominance may also be investigated as a function of speeds. This is shown in Figure \ref{f4}a--c as the number distribution of each type of motion with speed and position distinguished.
The Doppler speed distribution in Figure \ref{f4}a exhibits a peaked profile, similar to that observed for NBPs (Figure \ref{f4}c). This similarity suggests that both phenomena may be driven by mechanisms operating at characteristic scales. While NBPs are known to be driven by granule-scale magnetoconvection, the driving mechanism for spicules remains unclear. The speed distribution of blueshifted spicules has a peak at lower speeds (pink arrow) compared to typical RBE velocities. We propose that \RBE\ initially emerge at low speeds and subsequently undergo acceleration. The varying acceleration efficiency among the spicules naturally produces the observed decreasing speed distribution towards higher velocities (Figure \ref{f4}a).

On the other hand, {\RRE} appear in a speed distribution that monotonically decreases from zero speed (Figure \ref{f4}b). This implies a straightforward cascade that is probably governed by free falling under gravity. 
The dotted guidelines in Figure \ref{f4}ab show that sometimes the total size of {\RRE} can be comparable to that of {\RBE} in some speed range. This is a phenomenon already seen in Figure \ref{f2}d that one redshifted spicule occupies a wider area at a high speed, so as to match the {\RBE} in size when we limit the comparison to a particular speed range. If these numbers are a proxy for mass, this could be mean that uplifted materials make an episodal falling to a preferred location (region B) en mass at a high speed attained by gravity. 
More importantly, NBPs show a strong regional variation such that more NBPs reside in region B than in A or C. Blueshifted spicules  show no obvious regional variation, although they form a speed distribution similar to that of NBPs. Redshifted spicules tend to be more concentrated in A, but we regard it as an episodal behavior. Spicules do not seem to be concentration in region B not because they are infrequent there, but equally frequent in all regions.

For the spicules exhibiting torsional motion, the Doppler speeds of spicules can be meaningfully compared with the NBP velocities. Since they are mostly of transversal motions, their ratio may be related to the density ratio under the energy conservation.
The peak velocity of \RBE\ is found near 9 \kms\ only a little larger than that of NBPs,  0.7 {\kms}. This is not surprising because at their bases they are similar and later amplified upon propagation. For a fair comparison, we utilize the published results for the torsional motion of 25--30 \kms\ \citep{DePontieu12} to find the ratio $ (v_{\rm sp}/v_{\rm NBP})^2$ as high as (0.46--1.8)$\times 10^3$.
On the other hand, the expected density contrast is in the range of $2\times 10^{5}$ to $2\times 10^{7}$, considering the mass density in the photosphere  $2\times 10^{-7}$ g cm$^{-3}$  obtained from the Global Oscillation Network Group (GONG) project \citep{Dalsgaard1996Sci...272.1286C} and spicule density varying between $10^{-12}$--$10^{-14}$ g cm$^{-3}$ \citep{Goodman2012}, $3\times 10^{-13}$ g cm$^{-3}$ \citep{2000Sterling} and $10^{-13}$ g cm$^{-3}$ \citep{suematsu+wang+zirin1995}. Therefore our estimated density ratio is much smaller than expected. If we instead use the total speeds from \citet{DePontieu12} \sm60--110 {\kms}, the ratio comes out as $7.3\times 10^3 \leq (v_{\rm sp}/v_{\rm NBP})^2 \leq 2.5\times 10^5$ closer to the expected density ratio, but still lower. We tentatively assume that only a portion of the NBP energy is used for the torsional motion of spicules and the rest goes to their field-aligned motion.

\begin{figure}[tbh]  
\includegraphics[width=\textwidth]{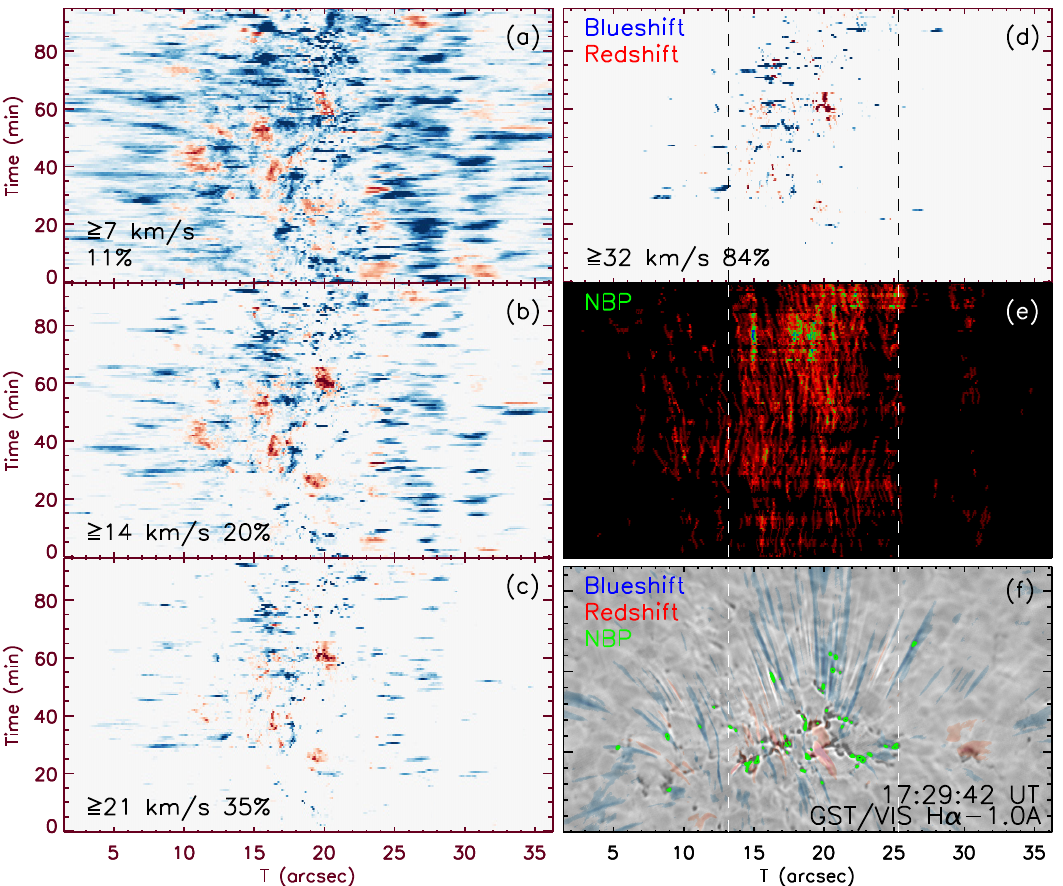}
\caption{TD maps for spicules and NBPs. (a--d) TD maps for spicules calculated for the different speed threshold as denoted in each panel. See text for the construction method. (e) TD map for NBPs obtained by adding up the number of pixels occupied by NBPs along the vertical direction ($R$-axis) to result in the one-dimensional distribution in the $T$-axis at each time.  
(f) {\RBE}/{\RRE} shown as color shades over the {\ha}--1.0 {\AA} inverse image as a guide.  
The vertical dashed lines divide the FOV into three regions as defined in previous figures.}
\label{f5}
\end{figure}

\subsection{TD Maps for Spicules and NBPs}\label{sec:td}

Since both spicules and NBPs are highly transient phenomena, their characteristic properties need to be determined through statistical analysis over sufficiently long time periods. 
Time–distance (TD) maps provide a statistical means of displaying spatiotemporal variations on a two-dimensional space. One limitation of this method is that the same spicules may be counted multiple times. To construct the TD map, we placed a slit oriented as close as possible to perpendicular to the spicule axis. Given the 40 s cadence of our observations, a spicule moving at 100 \kms\ would be displaced by approximately 4,000 km between consecutive frames. Consequently, spicules shorter than this distance are likely to be detected only once.
The TD maps presented in Figure \ref{f5} are constructed in this alternative approach of setting the slit perpendicular to the axes of spicules. 
Specifically, for blue/redshifted spicules (Figure \ref{f5}a--d), we construct the TD maps by counting at each time and $T$-position, all pixels along the $R$-axis that have Doppler speeds exceeding the threshold indicated in each panel. The resulting ratio of the total \RBE\ to \RRE\ count is also shown.
A single criterion is adopted for NBPs, the intensity contrast above 7\%, as the result is less sensitive to this criterion (Figure \ref{f5}e).

Figure \ref{f5}f shows the \RBE\ (blue color shade), \RRE\ (red), and NBPs (green) overlaid on an {\ha} far wing image along with the long-dashed lines denoting our three sections, A--C, as defined in the previous figures. NBPs are more concentrated in region B which coincides with the EUV bright region in the SDO/AIA 193 {\AA} image (see Figure \ref{f1}).
The \RRE\ tend to appear in patches while the \RBE\ are widely spread especially at lower speeds--thus does not show correlation with the coronal emission (see Figure \ref{f4}). A rather surprising result is that the higher speed spicules the more concentrated in region B, taking after that of NBPs. The observation that the fastest spicules and NBPs have similar spatio-temporal distribution is remarkable considering the very different nature of the spicule speed and NBP density. This suggests a potential physical link between photospheric NBPs and chromospheric spicule acceleration \citep[cf., ][]{bose2021a}. 

The relative dominance of \RBE\ over {\RRE} quantified by their areal ratio in images exhibits strong speed dependence.  
When considering the full Doppler velocity range, \RRE\ constitute merely 10\% of {\RBE}, reflecting predominantly upward plasma motion. This ratio increases steadily with speed, as shown in Figure \ref{f5}a--d, reaching 90\% at high speeds $\geq$35 {\kms}. 
These area-based ratios differ from the previous count ratio of RBEs to RREs, as individual RBEs and RREs can occupy markedly different areas. 
Yet the dominance of blueshifted to redshifted spicules is in overall agreement to the RBE-RRE count ratio reported by \citet{Sekse+etal2013ApJ...769...44S}. 
The newly found tendency of balance toward higher speeds is partly due to the redshifted spicules at lower height having gained high speed under their gravitational fall. There are also contributions from finely structured alternating pattern of blue- and redshifts in close distances in region B.
The trend of strong blue- and redshifted spicules being balanced in size toward higher speed may bear implications on the driving mechanism.

\subsection{Correlation with the EUV Brightness Distribution}

\begin{figure}[tbh]  
\includegraphics[width=\textwidth]{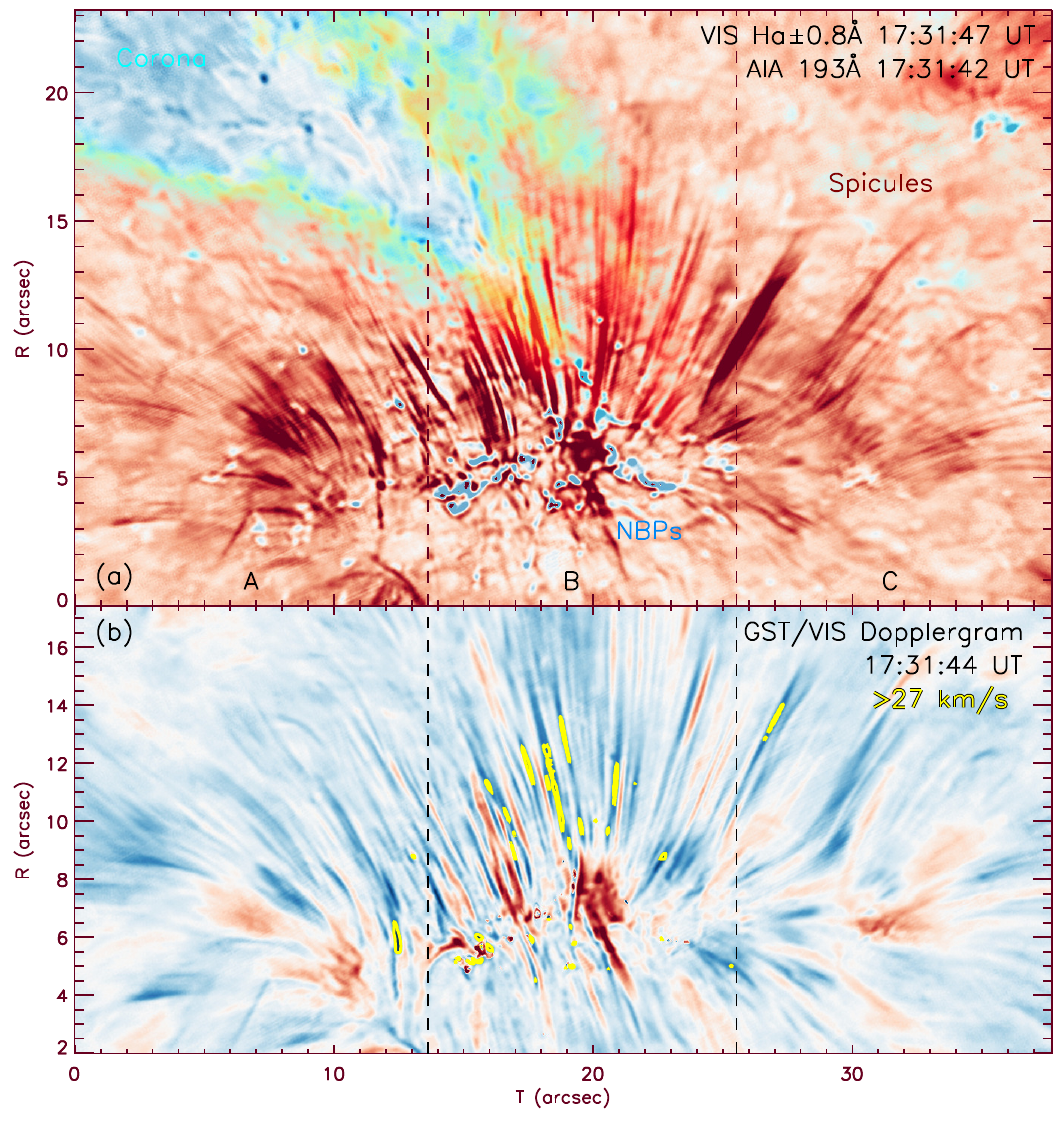}
\caption{Coronal, chromospheric, and photospheric activities. (a) AIA 194 {\AA} intensity overlaid as color shades over a composite GST/VIS {\ha} image where the red and blue wing $\pm$1.0 {\AA} images added. 
Spicules in both wings appear as dark narrow features and NBPs appear bright. 
(b) The {\ha} Dopplergram showing the blueshifted/redshifted spicules in blue/red colors. The yellow contours mark high speed ($\geq$ 27 {\kms}) regions.
The vertical dashed lines divide the FOV into the three regions as in the previous figures. 
An animation is available showing this figure with grayscale background in the top for the points A from 16:31:40 UT to 17:56:42 UT.
}
\label{f6}
\end{figure}

Finally we compare the spatial distributions of spicules and NBPs against that of the EUV brightness.
While the TD maps offer useful proxies for spicule distributions, their interpretation is limited by the 3D nature of spicules, which extend in various orientations.  
Visual inspection of spicules alongside coronal emission in the images may provide additional insight.

Figure \ref{f6}a shows the SDO/AIA 193 {\AA} intensity map as color shades overlaid over the GST/VIS {\ha} composite image, i.e., two wings $\pm$0.8 {\AA} images added together as in Figure \ref{f1}. The SDO/AIA 193 \AA\ image is chosen because the coronal hole and the loop structure are best visible in this channel. However, it may not be fully representative of the corona in that the peak temperature response of the 193 \AA\ is at $\log T \approx 6.2$ with substantial contamination from cooler transition region (TR) ions at $<$500,000 K 
\citep{ODwyer2010A&A...521A..21O, Martinez-Sykora2011ApJ...743...23M, DelZanna2011A&A...535A..46D, Bose2025ApJ...983L...7B} which could imply heating to TR temperatures more than coronal temperatures, but still significantly higher than chromospheric tempperatures.
The dark red features are spicules and the bright blue features are NBPs so that these three components represent activities in the corona, chromosphere, and photosphere, respectively.
Here we confirm that NBPs show stronger concentration in region B, correlating more closely with the EUV corona, while spicules exhibit no clear spatial association with coronal features as they occur rather uniformly across all sections. 

Figure \ref{f6}b shows the Doppler map constructed using the 11 wavelength point {\ha} images between $\pm$1.0 {\AA}. 
Both blueshifted and redshifted spicules are displayed as color scale images together and the superimposed yellow contours outline the regions of Doppler speeds above 27 {\kms}. Those high-speed spicules appear more concentrated—though not exclusively—in region B, while low-speed blueshifted/{\RRE} are found in all sections.
Another key characteristic linking the spicules to the coronal emission is the alternating pattern of thin {\RBE} and {\RRE} in short intervals predominantly found in B. 
We thus suggest that high speeds, alternating pattern of thin blueshifted and {\RRE} determine whether spicules are correlated with coronal emission.
To compare with the result of \citet{Nived_2022} 
one of these features that most closely matches their criterion for the spicule-corona connection (see  \S \ref{sec:intro}) is the alternating pattern of thin {\RBE}/{\RRE}, as they are narrowly spaced to be densely packed in a small space, resulting in a high occurrence rate.
The comparison between the \ha\ image–derived quantities and the coronal emission suggests that photospheric motions, as represented by the NBPs, could serve as a major source of upward energy transfer.
 
\section{Discussions and Conclusion}

The properties of spicules and their potential contribution to coronal heating have been extensively studied over the past decades. Nevertheless, the present work aims to complement previous analyses of the spicule–corona connection by incorporating NBPs into the picture. Our focus lies in examining the dynamics and statistical properties of spicules using TD maps and speed distributions, and those of NBPs through automated feature tracking with the SWAMIS algorithm applied to high-resolution \ha\ observations from the BBSO GST/VIS instrument.
With Dopplergrams, the torsional components of spicule motion cannot be easily separated from the swaying components, particularly when their projected velocities are comparable \citep{Sekse+etal2013ApJ...769...44S}. 
However, given the typical differences in characteristic speeds among field-aligned, torsional, and swaying motions \citep{DePontieu12}, we could infer, from their velocity dependence, additional information on  the coupling between photospheric activity and chromospheric--coronal activities. We summarize our main findings as follows.

\begin{enumerate}

\item 
For the first time, we incorporated NBP motions in the analyses of the spicule--corona relationship to find a good correlation among NBPs, high-speed and narrow spicules, and the EUV brightness. 
Our result that not all but a subset of spicules is aligned with the coronal brightness  is resonant with the result that only spicules above some threshold directly contribute to the coronal heating \citep{Nived_2022}. 

\item 
Alternating signs of Dopplershifts of spicules interpreted as torsional motion is faster than the NBP velocity by a factor of 10--$10^2$, which may related to the mass density ratio in two different heights as $10^2$--$10^4$ if rigid body rotation is assumed.

\item
More \RBE\ than {\RRE} are found in general, but the relative dominance of \RBE\ over \RRE\ reduces with increasing Doppler speeds.
If we include all the range of Doppler speeds $\geq$4 {\kms}, {\RRE} constitute only 10\% of {\RBE}, but increases steadily with velocity, reaching 90\% at high speeds $\geq$35 {\kms}. 

\item 
The speed distributions of \RBE\ and NBPs both exhibit distinct peaks, consistent with the characteristics of driven processes. In contrast, \RRE\ display a monotonically decreasing speed distribution starting from zero velocity. Together with the height difference between the blue and redshifted spicules, this result indicates significant gravitational influence.

\item 
A population of \RRE\ lying at greater heights share similar lengths, widths, and speeds with {\RBE}. The morphological and dynamical similarity between these \RRE\ and \RBE\ implies the same driving mechanism operating for them. 

\item 
NBPs are concentrated under the  AIA 193 \AA\ bright region, while spicules are ubiquitous in general, except the fastest spicules tending to form above NBPs.
The high-speed spicules preferentially occuring in the NBP-dominated region is a more specific association than the known link between RBEs/RREs and general network fields. 

\end{enumerate}

Subject to the projection effects and the ambiguity about the field aligned vs transverse motions, these properties of \RBE\ relative to those of \RRE\ do not directly translated to the up and downflows.  
Nonetheless, the trend of their balance toward high Doppler speeds seems to imply that the driving mechanism should provide thrust in both directions along or within narrow channels.
Spicules driven upwardly by reconnection in the lower atmosphere along a thin flux tube may satisfy some of these constraints, but not the downflowing spicules. 
For the downward mass motions, the low-lying redshifted spicules can largely be attributed to gravitational fallback. The high-lying redshifted spicules needs alternative drive. A possibility is the impulsive reconnection events occurring at the interface between the upper TR and the lower corona, as proposed in 3D MHD simulations \citep{Hansteen2010ApJ...718.1070H} supported by observations in the TR and chromosphere \citep{McIntosh2012ApJ...749...60M, bose2021b}.

NBPs provides an alternative possibility. Associated with convection in open fields, NBPs' motion can generate upwardly propagating \alf\ waves. However, \alf\ waves are known to suffer significant reflection in the TR \citep{Zhugzhda1982, Cranmer2005}. The {\alf} waves propagating upward and reflected downward can impart a portion of their energy to the chromospheric spicules bidirectionally, which can explain why \RRE\ and \RBE\ share similar properties including speed, length, and thickness. The wave hypothesis is also compatible with the existing interpretation that adjacent \RBE\ and \RRE\ represent torsional waves.
The observed clustering of NBPs under the coronal loop may imply enhancement in either generating the \alf\ waves 
propagating into the corona through the field lines \citep{Lee2025ApJ...988L..16L} or shuffling the field lines to cause topological dissipation as in the nanoflare model \citep{Parker1988}. 
The varying symmetry in low and high speeds still needs an explanation. Maybe spicules cannot strictly follow {\alf} waves due to either varying degree of coupling between the waves and plasma depending on the local degree of ionization \citep{MartinezSykora+etal2013ApJ...771...66M, MartinezSykora2017Sci...356.1269M, Martinez-Sykora2020ApJ...889...95M, Erdelyi2004}.
 
We conclude that NBPs are the main energy donor and waves are the main energy carrier and that spicules rather emerge as byproducts of {\alf} waves based on the locations and dynamics of spicules and NBPs. Future works should aim at distinguishing the torsional motions from the field-aligned motions with more precise Doppler measurements and higher cadence imaging. Observations of small-scale solar ejections with higher spectral resolution across multiple spectral lines in the chromosphere and photosphere are anticipated to play a key role, particularly with facilities such as DKIST and EST.


\begin{acknowledgments}
This work was supported by NSF grants, AGS-2114201, AGS-2229064 and AGS-2309939, and NASA grants, 80NSSC19K0257, 80NSSC20K0025, 80NSSC20K1282 and 80NSSC24K0258. E.-K.L. acknowledges support by the Korea Astronomy and Space Science Institute under the R \& D program of the Korean government (MSIT; No. 2025-1-850-02). BBSO operation is supported by a US NSF AGS 2309939 grant and the New Jersey Institute of Technology. The GST operation is partly supported by the Korea Astronomy and Space Science Institute and the Seoul National University. 
\end{acknowledgments}

\vspace{5mm}
\facilities{BBSO (GST/VIS and GST/NIRIS)}

\software{IDL, SolarSoft \citep{2012ascl.soft08013F}, SWAMIS \citep{DeForest_2007}, OF  \citep{2022XYang}}

\vspace{5mm}

\appendix

\section{Methods for Determining LOS speeds of Spicules}
\label{ap:1}

We discuss three methods for determining the Doppler speeds. One is more practical for constructing Doppler maps and the other two are  needed for magnitudes of high-speed UEs/DEs.
In the so-called center-of-gravity (COG) method, the Doppler shift  $\Delta \lambda_{\rm COG}$ and the equivalent width  $W_{\rm COG}$  are  determined using the residual intensity profile,  $\Delta I_\lambda = I_{\rm ref}-I_\lambda$, of a given line profile $I_\lambda$ to the reference profile $I_{\rm ref}$. They are expressed as:
\begin{equation} 
\lambda_{\rm COG} = \frac{\int (\lambda-\lambda_0) \Delta I_ \lambda d\lambda}{\int \Delta I_\lambda d\lambda}, \qquad
W_{\rm COG}  =  \Big[ \frac{\int (\lambda-\lambda_0)^2 \Delta I_\lambda d\lambda}{\int  \Delta I_\lambda d\lambda} \Big]^{1/2}
\end{equation}  
At least two types of $\lambda_{\rm COG}$ may result, depending on which profile is used for $I_{\rm ref}$. If we use the ambient continuum intensity as the reference, $I_{\rm ref}=I_{\rm con}$  \citep{Uitenbroek2003}, the core part has higher weights, resulting in low speeds. If we use the mean spectrum of the whole FOV $I_{\rm ref}= I_{\rm avg}$,  \citep{Rouppe+etal2009ApJ...705..272R}, more weight is given to the wing enhancement resulting in higher speeds.

\begin{figure}[tbh]  
\includegraphics[width=\textwidth]{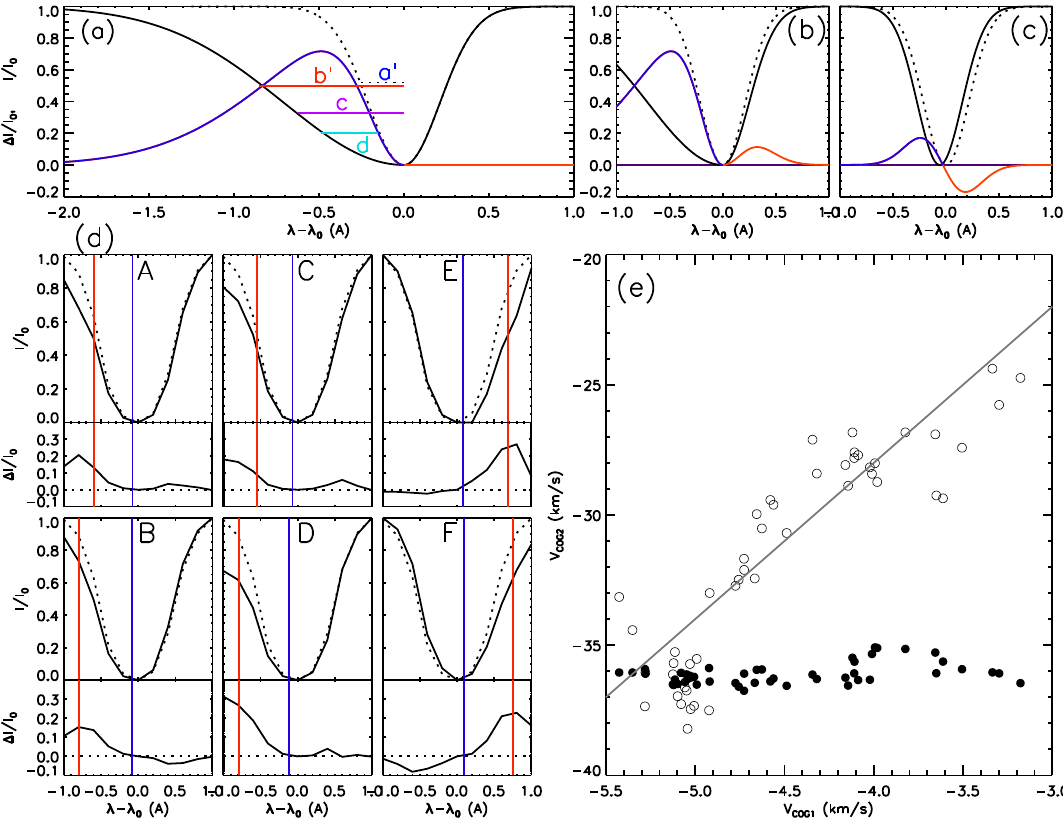}
\caption{Theoretical and observational line profiles. (a) An asymmetric line profile with Gaussians with scale length $b$ in the blue wing and $a$ in the red wing (solid black line) matching that of mean profile (dotted). $a'=a/\sqrt{2\ln 2}$~and $b'=b/\sqrt{2\ln 2}$ are the distances from the line center to the wavelengths of the half maximum intensity. $c=(a+b)/\sqrt{\pi}$ represents the characteristic wavelength of the residual profile (blue-red), and $d=(b-a)/\sqrt{\pi}$ is the wavelength difference representing the line asymmetry. Two other cases are shown with the residual intensity in the red wing being positive (b) and negative (c).
(d) Normalized {\ha} line profiles at six selected points as denoted by the cross symbols in Figure \ref{f_a2}. Each box shows  the local {\ha} line profile and the average profile in the top and the difference profile in the bottom. The vertical guide lines indicate the COG wavelengths determined by method 1 (blue) and method 2 (red). (e) A scatter plot of the Doppler speeds determined by methods 1 and 2 (filled circles) and by methods 1 and 3 (open circles).
}
\label{f_a1}
\end{figure}

A simple calculation can show that the results of both methods should be closely related to each other. In an example shown in Figure \ref{f_a1}a, A Gaussian theoretical line profile with width $a$ is used for the mean profile, $\phi_0$ (the dotted lines in a--c), and an asymmetric line profile consisting of the red wing same as the mean profile and the more extended blue wing $\phi$  with $b>a$ (solid lines).
In method 1, the integral is performed on both sides of the center wavelength $\lambda_0$ so that 
$\Delta\lambda_{\rm COG}= \int_{-\infty}^{+\infty} \lambda'(1-\phi) d\lambda'/\int_{-\infty}^{+\infty}  1-\phi~d\lambda' =-(b-a)/\sqrt{\pi}$. 
In method 2, the integral is performed on the blueward wavelengths only where the ``excursion feature’’ shows up as absorption. In this case, 
$\Delta \lambda_{\rm COG}=
\int_{-\infty}^0 \lambda'(\phi_0-\phi) d\lambda'/\int_{-\infty}^0 \phi_0-\phi~d\lambda'
=-(b+a)/\sqrt{\pi}$, which is longer than the above by $2a/\sqrt{\pi}$. 
Method 2 is therefore adequate for detecting high speeds of RBEs (or RREs), but does not take into account the profile in the other side of the wing.   We can also consider a third method in which the integral interval is extended to encompass both sides of the line wings, i.e., $\Delta \lambda_{\rm COG}'=
\int_{-\infty}^\infty \lambda'(\phi_0-\phi)d\lambda'/\int_{-\infty}^\infty (\phi_0-\phi)d\lambda'$ to fully count the line asymmetry. 

The results of these three methods will depend on the residuals in the wings. Other studies use the wing residuals as a threshold for qualifying spicules as RBEs/RREs, for instance, 0.1 \citep{Lim2025} or 0.4 \citep{Nived_2022}. We note that such a criterion may yield the speeds in a narrow range by limiting the amount of line asymmetry in method 2, as will  be shown in  Figure \ref{f_a1}e.
This does not change the result of the case shown in Figure \ref{f_a1}a with no residual, but does affect the results in two other cases are shown in Figure \ref{f_a1}bc.
The profile shown in Figure \ref{f_a1}b includes line broadening as well as blueward excursion, resulting in the positive residual intensity in the other side. The profile in Figure \ref{f_a1}c shows a case of the line profile shifting to one side as a whole without broadening,resulting in the negative residual intensity. Method 1 still gives the same result for all cases, but methods 2 and 3 yield differing results. In particular, method 3 may result in a diverging speed for exactly equal amounts of Doppler shifts in both wings, which can be avoided by calculating the Doppler speed in each wing separately, and take average of them.

\begin{figure}[tbh]  
\includegraphics[width=\textwidth]{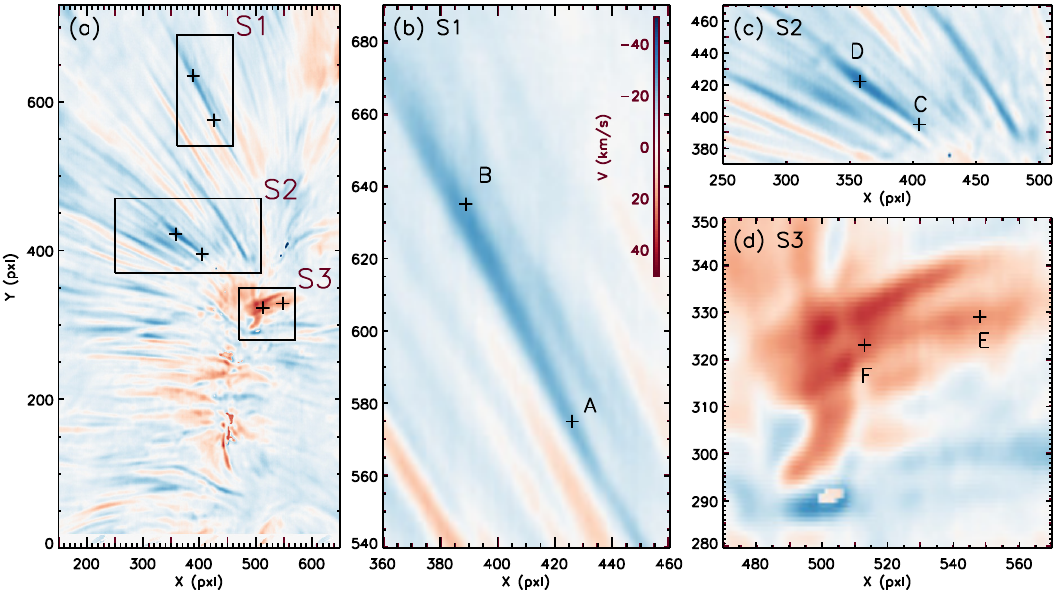}
\caption{A Dopplergram constructed from a set of GST/VIS {\ha} images at 11-wavelength points. (a) A larger FOV Dopplergram denoting three sub regions shown in (b-d). Blue/red color represent {\RBE} and {\RRE} and points A--F are selected to display local line profiles in Figure \ref{f_a1}. This map has been constructed using method 1 and the scale bar in (b) refers to the Doppler speeds determined using method 3. 
}
\label{f_a2}
\end{figure}

The six panels of Figure \ref{f_a1}d are meant to check the results obtained by applying methods 1 and 2 to the GST/VIS data. 
Panels in Figure \ref{f_a1}d show the local line profiles from the six selected points, A--F, as denoted in the Doppler map Figure \ref{f_a2}. 
Each panel shows the normalized local line profile (solid line), our average line profile (dotted), and their difference profile (solid line in the bottom). The vertical lines denote $\Delta \lambda_{\rm COG}$ determined by method 1 (blue) and method 2 (red) if lying within the spectral window $-1$ {\AA} $\leq \lambda' \leq 1$ {\AA}.
Whereas the differences of the method 1 results are so small to be noticeable, those of the method 2 results are visible more clearly.
The first and second column shows spectra at both ends of two {\RBE}s showing the characteristics of RBEs, and the LOS velocity difference between A and B, and that between C and D implies an upward acceleration. 
Likewise the two spectra at E and F of an {\RRE} in the third column shows the downward acceleration. 
Figure \ref{f_a1}e shows a scatter plot of the Doppler speeds determined by methods 1--3. The results obtained using method 2 lie in a narrow range after the criterion set for RBEs/RREs is applied and the integration performed only in one wing side. 
On the other hand, the results of method 3 obtained by performing the integral of in both wings, show a good linear correlation with those of method 1, represented by the relation $V_3 = 6V_1 + 4$ {\kms}, where $V_1$ and $V_3$ are the outputs of method 1 and 3, respectively, in units of {\kms}. We thus regard method 3 as best serving for our purpose.

Spatial locations of the local line profiles spectra are shown in Figure \ref{f_a2}. Here spicule locations are determined using method 1, and the local Doppler speeds in those locations are calculated using method 3. Method 1 produces a stable Doppler map at the cost of denouncing the high Doppler speeds, whereas method 3 yields a wilder map due to fluctuations in the line wings. The upward acceleration of {\RBE}s can be inferred from the local spectra (Figure \ref{f_a1}d) for the pairs, A--B and C--D (Figure \ref{f_a2}bc), and the downward acceleration from those of E-F (Figure \ref{f_a2}d).
In particular, B shows line profile shifting to the blue wing as a whole, meaning that even within a spicule, some parts may not show the characteristic of RBEs, while most other parts do.  
These results demonstrate that our Dopplergram can detect not only high speeds of RBEs and RREs, but also those of other spicules without the characteristics of RBE/RREs in a consistent manner.

\bibliographystyle{aasjournal}
\end{document}